\documentclass[12pt]{article}
\usepackage{fancyhdr}

\usepackage{overcite}

\usepackage{graphicx}

\usepackage{nicefrac}

\usepackage{amsfonts}

\usepackage{amsmath}

\usepackage{textgreek}

\usepackage{booktabs}

\usepackage{epstopdf}

\usepackage{caption}

\usepackage{subcaption}

\usepackage{authblk}

\usepackage{setspace}

\usepackage{multirow}

\usepackage{makecell}

\usepackage{rotating}

\makeatletter \renewcommand\@biblabel[1]{$^{#1}$} \makeatother
\newlength{\bibhang}
\title{Introducing EGSnrc application egs\_mird}
\author{Martin P. Martinov}
\author{Chidera Opera}
\author{Rowan M. Thomson}
\author{Ting-Yim Lee}

\newcommand{\ie}{{\it i.e.}, }
\newcommand{\eg}{{\it e.g.}, }
\newcommand{\etal}{{\it et al}}

\newcommand{\cen}[1]{\begin{center} #1 \end{center}}

\newcommand{\captionl}[2]{\parbox{16.5cm}{\caption[#1]{{\sf #2}}}}

\newcommand{\captionlw}[2]{\parbox{21.85cm}{\caption[#1]{{\sf #2}}}}

\newcommand{\mm} {mm }
\newcommand{\cm} {cm }

\newcommand{\mms}{mm}
\newcommand{\cms}{cm}

\newcommand{\uGy} {$\mu$Gy$\cdot$GBq$^{-1}$$\cdot$s$^{-1}$ }
\newcommand{\uGys}{$\mu$Gy$\cdot$GBq$^{-1}$$\cdot$s$^{-1}$}

\newcommand{\dens}{g$\cdot$cm$^{-3}$}

\begin{document}

\cen{\sf {\Large {\bfseries Fast full patient and radioisotope Monte Carlo simulations of targeted radionuclide therapy: introducing egs\_mird} \\
Martin P. Martinov$^\text{a)}$, Chidera Opera, Rowan M. Thomson, and Ting-Yim Lee$^\text{b)}$} \\ \vspace{3mm}
email: a) martinov@physics.carleton.ca and b) tlee@robarts.ca \\ \vspace{5mm}
Robarts Research Institute, London, Ontario, N6A 5K8, Canada \\ \vspace{5mm}
Lawson Health Research Institute, London, Ontario, N6C 2R5, Canada \\ \vspace{5mm}
Carleton Laboratory for Radiotherapy Physics, Department of Physics, Carleton University, Ottawa, Ontario, K1S 5B6, Canada \\ \vspace{5mm} 
submitted on \today \\
}

\pagestyle{empty}
\begin{abstract}
{\small
\noindent\textbf{Background}: Targeted Radionuclide Therapy (TRT) is a fast-growing field garnering much interest, with several clinical trials currently underway, that has a steady increase in development of treatment techniques.  Unfortunately, within the field and many clinical trials, the dosimetry calculation techniques used remain relatively simple, often using a mix of S-value calculations and kernel convolutions.

\noindent\textbf{Purpose}: The common TRT calculation techniques, although very quick, can often ignore important aspects of patient anatomy and radionuclide distribution, as well as the interplay there-in.  This paper introduces egs\_mird, a new Monte Carlo (MC) application built in EGSnrc which allows users to model full patient tissue and density (using clinical CT images) and radionuclide distribution (using clinical PET images) for fast and detailed dose TRT calculation.

\noindent\textbf{Methods}: The novel application egs\_mird is introduced along with a general outline of the structure of egs\_mird simulations.  The general structure of the code, and the track-length estimator scoring implementation for variance reduction, is described.  A new egs++ source class egs\_internal\_source, created to allow detailed patient-wide source distribution, and a modified version of egs\_radionuclide\_source, changed to be able to work with egs\_internal\_source, are also described.  The new code is compared to other MC calculations of S-values kernels of $^{131}$I, $^{90}$Y, and $^{177}$Lu in the literature, along with further self-validation using a histogram variant of the electron Fano test.  Several full patient prostate $^{177}$Lu TRT prostate cancer treatment simulations are performed using a single set of patient DICOM CT and [$^{18}$F]-DCFPyL PET data.

\noindent\textbf{Results}: Good agreement is found between S-value kernels calculated using egs\_mird with egs\_internal\_source and those found in the literature.  Calculating 1000 doses (individual voxel uncertainties of $\sim$0.05\%) in a voxel grid Fano test for monoenergetic 500~keV electrons and $^{177}$Lu electrons results in 94 and 99\% of the doses being within 0.1\% of the expected dose, respectively.  Patient prostate, rectum, bone marrow, and bladder dose volume histograms (DVHs) results did not vary significantly when using the track-length estimator and not modelling electron transport, modelling bone marrow explicitly (rather than using generic tissue compositions), and reducing activity to voxels containing partial or full calcifications to half or none, respectively.  Dose profiles through different regions demonstrate there are some differences with model choices not seen in the DVH.  Simulations using the track-length estimator can be completed in under 15 minutes ($\sim$30 minutes when using standard interaction scoring).

\noindent\textbf{Conclusion}: This work shows egs\_mird to be a reliable MC code for computing TRT doses as realistically as the patient CT and PET data allow.  Furthermore, the code can compute doses to sub-1\% uncertainty within 15 minutes, with little to no optimization.  Thus, this work supports the use of egs\_mird for dose calculations in TRT.
}
\end{abstract}

\setlength{\baselineskip}{0.7cm}
\pagestyle{fancy}

\section{Introduction} \label{sec:Introduction}

The novel and growing field of Targeted Radionuclide Therapy (TRT) dosimetry still relies heavily on calculations performed using kernel convolution or S-value computations\cite{Mo20}, compared to more detailed calculations such as Monte Carlo (MC).  These typical calculation methods, although reliant on assumptions of patient anatomy or activity distribution, offer results in very little time, a requirement for any type of clinical implementation.  Towards the goal of fast and more detailed dose calculations, this paper introduces egs\_mird, a MC application built in EGSnrc\cite{Ka99a} capable of full simulations of TRT patients with electron and photon radionuclide activity.  Using a photon track-length estimator and custom sources, egs\_mird can calculate full DICOM RT dose distributions from DICOM PET and CT data (converted to EGSnrc format using in-house tools) in under 15 minutes on an AMD Ryzen 9 3900XT 12-core CPU.

TRT is an exciting new treatment modality with very high potential for dose localization to the target compared to most typical radiotherapy treatments.  As of publication, there are several phase II clinical trials underway for TRT treatments of prostate cancer treatments with $^{177}$Lu\cite{Vo20,Ho21}, which primarily use DPK convolution to calculate treatment dose\cite{Vo19}.  Using MC, dose calculations can be made at the resolution of the CT data using full patient media and density information with a model that includes the fully heterogeneous activity distribution read in using PET data.  Although MC has been used in this field previously, it is often prohibited by assumptions in the model\cite{Sa14} or by the total computation time\cite{Sa17}.  The purpose of egs\_mird is to provide dose results which are both fast and reliable in a model containing detailed patient anatomy such as bone, organs-at-risk, and even calcifications in the treatment volume.

As egs\_mird makes use of the egs++ egs\_radionuclide\_source\cite{To18}, it is capable of modelling the decay (or decay chains) of many radionuclides that would see use in TRT (\eg $^{32}$P, $^{89}$Sr, $^{90}$Y, $^{131}$I, $^{153}$Sm, $^{177}$Lu, $^{186}$Re, $^{188}$Re, $^{223}$Ra, and $^{225}$Ac\cite{Bi14}).  As EGSnrc cannot model alpha particle transport, all alpha particles generated in a decay are assumed to deposit their energy locally.  Though this could be a potential shortcoming on short length scales, for millimeter-sized voxels based on patient CT data, assuming local energy deposition should be a sufficient approximation considering alpha particle range\cite{Bi14}; similar to the no electron transport assumption made in this work when performing track-length simulations.  Though this work simulates $^{131}$I, $^{90}$Y, and $^{177}$Lu for data validation, the patient dose calculations are only performed using $^{177}$Lu.  The goal is to investigate the effects of different transport parameters and options available in egs\_mird, as well as different modeling approaches to importing DICOM data.

This paper will provide a full introduction of egs\_mird and its associated codes.  It will discuss the specifics of the implementations using the EGSnrc code, demonstrate validity of the new code using a Fano test and cross-comparison with the literature, demonstrate the ability and speed using patient CT and PET data, and examine the final patient dose results.

\section{Methods} \label{sec:Methods}

EGSnrc commit 09aa30e of the EGSnrc GitHub master branch (specifically the egs++ package\cite{Ka05a}) is used as the base Monte Carlo code for all simulations in this work.  In addition, egs\_glib geometry from the CLRP GitHub page for EGSnrc\_with\_egs\_brachy commit 2718117 are incorporated into the installation\cite{Mc16} to model patient geometry using the egsphant format.  A modified version of egs\_radionuclide\_source\cite{To18} is used which allows for more flexible source inputs (such as egs\_internal\_source described below); the modified egs\_radionuclide\_source\cite{To18} is currently available as a pull request on the EGSnrc GitHub page, with plans to have it incorporated into the next annual release.  Electrons and photons, when simulated, are transported down to 1~keV kinetic energy.  Most MC transport parameters are the default EGSnrc settings, except as noted here.  Bound Compton scattering, Rayleigh scattering, and electron impact ionization are set to on, pair angular sampling is set to off, and brems cross sections are set to `nist'.  All media used in the simulations below are listed in Table~\ref{tab:mediaComp}, based primarily on ICRU46 compositions\cite{ICRU46} with some additional compositions used from the literature for marrow composition\cite{Sc00b}, water\cite{Ri04}, and air\cite{Ri04}.

\begin{table}[htbp]
	\vspace{0.8\normalbaselineskip}
	\centering
	\rotatebox{90}{
		\captionlw{Media Compositions}{The density and atomic composition of all elements used in this work with their accompanying source in the literature.  
		\label{tab:mediaComp}}}
	\raisebox{.5\textheight}[0pt][0pt]{\rotatebox[origin=c]{90}{
		\centering
		\footnotesize
		\begin{tabular}{llrrrrrrrrrrrl}
		\toprule
		\multicolumn{1}{c}{\multirow{2}[4]{*}{Medium}} & \multicolumn{1}{c}{\multirow{2}[4]{*}{\makecell{Density\\(\dens)}}} & \multicolumn{11}{c}{Atomic Mass Fractions}                                            & \multicolumn{1}{c}{\multirow{2}[4]{*}{Source Paper}} \\
		\cmidrule{3-13}          &       & \multicolumn{1}{c}{H} & \multicolumn{1}{c}{C} & \multicolumn{1}{c}{N} & \multicolumn{1}{c}{O} & \multicolumn{1}{c}{Na} & \multicolumn{1}{c}{P} & \multicolumn{1}{c}{S} & \multicolumn{1}{c}{Cl} & \multicolumn{1}{c}{K} & \multicolumn{1}{c}{Ca} & \multicolumn{1}{c}{Ar} &  \\
		\midrule
		Male Tissue 								& 1.03 & 10.50 & 25.60 & 2.700 & 60.20 & 0.100 & 0.200 & 0.300 & 0.200 & 0.200 &       &       & ICRU46\cite{ICRU46} \\
		Cortical Bone 								& 1.92 & 3.40  & 15.50 & 4.200 & 43.50 & 0.100 & 0.200 & 0.300 & 0.200 & 0.200 &       &       & ICRU46\cite{ICRU46} \\
		Pros.	 									& 1.04 & 10.50 & 8.900 & 2.500 & 77.40 & 0.200 & 0.100 & 0.200 & 0.200 & 0.000 & 0.000 &       & ICRU46\cite{ICRU46} \\
		Calc.		 								& 3.06 & 0.30  & 1.600 & 0.500 & 40.70 & 0.000 & 18.70 & 0.000 & 0.000 & 0.000 & 38.20 &       & ICRU46 (for breast)\cite{ICRU46} \\
		Pros. \& Calc.								& 1.55 & 5.40  & 5.250 & 1.500 & 59.05 & 0.100 & 9.400 & 0.100 & 0.000 & 0.100 & 19.10 &       & 50/50 hybrid medium \\
		Rectum 										& 0.75 & 6.30  & 12.10 & 2.200 & 78.80 & 0.010 & 0.100 & 0.000 & 0.100 & 0.100 & 0.000 &       & ICRU46\cite{ICRU46} \\
		Bladder (full) 								& 1.03 & 10.80 & 3.500 & 1.500 & 83.00 & 0.300 & 0.100 & 0.100 & 0.500 & 0.200 & 0.000 &       & ICRU46\cite{ICRU46} \\
		Marrow Yellow 								& 0.98 & 11.50 & 64.60 & 0.700 & 23.20 &       &       &       &       &       &       &       & Schneider \textit{et al}\cite{Sc00b} \\
		Marrow Y. \& R. 							& 1.00 & 11.00 & 53.10 & 2.100 & 33.60 &       & 0.100 &       &       &       &       &       & Schneider \textit{et al}\cite{Sc00b} \\
		Marrow Red 									& 1.03 & 10.60 & 41.70 & 3.400 & 44.20 &       & 0.100 &       &       &       &       &       & Schneider \textit{et al}\cite{Sc00b} \\
		Air*   										& 1.20 & 0.07  & 0.010 & 75.03 & 23.61 &       &       &       &       &       &       & 1.270 & Rivard \textit{et al}\cite{Ri04} \\
		Water 										& 1.00 & 11.11 &       &       & 89.99 &       &       &       &       &       &       &       & Rivard \textit{et al}\cite{Ri04} \\
		Tissue (DPK)\textsuperscript{\textdagger}	& 1.04 & 10.45 & 22.66 & 2.490 & 63.53 & 0.112 & 0.134 & 0.204 & 0.133 & 0.208 & 0.024 &       & Cristy and Eckerman\cite{Cr87} \\
		\bottomrule \vspace{-0.6\normalbaselineskip} \\
		\multicolumn{14}{l}{* Air density is in units of mg$\cdot$cm$^{-3}$.} \\ 
		\multicolumn{14}{l}{\textsuperscript{\textdagger} Tissue (DPK) also includes trace amounts of Mg, Si, Fe, Zn, Rb, Zr, and Pb not listed on the table.} \\
		\end{tabular}}}
\end{table}

\subsection{Implementations in egs\_mird and egs\_internal\_source} \label{sec:Methods-code}

On a basic level, the application egs\_mird inherits egs\_advanced\_application with extended functionality to score dose in a 3d array and output an EGSnrc .3ddose file (or a binary .b3ddose file).  Additionally, it includes the track-length (TL) estimation variance reduction technique, similar to that of egs\_brachy\cite{Mc16}.  At its most bare bones implementation, egs\_mird takes in material, run control, geometry, and source definitions like all other egs++ applications.  It only requires one unique input block, `scoring options', in which the user specifies the geometry to score dose in, with the option to change the output filetype and filename, if desired.

Additionally, egs\_mird has an optional input block, `variance reduction', which allows the user to turn on track-length scoring and select a data file containing the appropriate mass-energy absorption coefficient data used for the calculation.  A photon TL estimator\cite{Wi87b} is a variance reduction technique which operates on the assumption that, at low photon energies, all electrons deposit their energy locally, thus, dose can be approximated as collision kerma.  It requires that the electron transport cutoff be set to the highest energy, \ie no electrons are transported in the simulation, they simply deposit their energy locally when generated.  When using TL scoring, a photon travelling through a volume is tallied as a dose deposition event.  The dose deposited in this event ($D_e$) to the scoring volume ($V$) can be calculated using the total track-length through the volume ($t$), the appropriate mass-energy absorption coefficient ($\nicefrac{\text{\textmu}_\text{en}}{\text{\textrho}}$), and the energy of the travelling photon ($E$) using the following equation:

\begin{equation}
	D_e = \frac{t \cdot E \cdot \nicefrac{\text{\textmu}_\text{en}}{\text{\textrho}}}{V} \label{eq:TLE}
\end{equation} \vspace{-0.8\normalbaselineskip}

Scoring dose along the track of a photon rather than in the individual interaction events allows for a very large increase in efficiency of simulations; even though a single photon history using TL scoring takes more time due to the above computations, the total number of histories required to achieve low uncertainties can be largely reduced.\cite{Ch16}

EGSnrc's egs++ package has many different options for generating source particles in a simulation, but all of them are limited by the fact they are generated using rudimentary shapes, and cannot efficiently reflect the simulation geometry.  Thus, a new source, egs\_internal\_source, is developed which takes two inputs: a rectilinear patient geometry (\ie patient defined as a 3D grid of voxels with appropriate media and density) and a list of voxel regions with corresponding Time-Integrated Activity (TIA).  Using the alias method\cite{KR79a}, already implemented in EGSnrc as the class EGS\_AliasTable, the list of regions and weight is built into a table that can be sampled at O(1) efficiency (\ie EGS\_AliasTable samples at the same speed, no matter how large the table is) using two random numbers to determine where to generate the particle.  Thus, if a user has a heterogeneous radionuclide TIA distribution over a patient geometry, they can incorporate it into a patient simulation.

The egs\_mird application and egs\_internal\_source for generating particles throughout a phantom are available at the https://github.com/Robarts-Lee-Lab/egs\_mird website.  The modified egs\_radionuclide\_source which allows any other source as a input (now able to be combined with egs\_internal\_source) is available as a pull request on the main EGSnrc github page at https://github.com/nrc-cnrc/EGSnrc.

\subsection{Fano test using egs\_mird and egs\_radionuclide\_source} \label{sec:Methods-validation}

Although egs\_mird and egs\_internal\_source rely quite heavily on code already within EGSnrc, as brand new implementations, they require validation.  Additionally, egs\_radionuclide\_source\cite{To18} is a relatively new feature in EGSnrc that has not seen use as extensively as many of the older elements of EGSnrc and, thus, should also be validated.  Towards the goal of validation, a modified Fano cavity test\cite{MT20,Bo15} is performed using egs\_mird, egs\_internal\_source, and egs\_radionuclide\_source.  A cubic scoring grid consisting of 10x10x10 cubic voxels, a (10~\cms)$^3$ volume, is placed at the center of a (20~\cms)$^3$ water phantom, allow a 5~\cm non-scoring buffer on every side.  All voxels are assigned water with a density inversely proportional to the distance from the center of the phantom to their respective midpoints to ensure heterogeneous mass distribution.  An egs\_internal\_source is generated with voxel weight (probability of generating source particle in the voxel) set to voxel density using egs\_internal\_source, creating a simulation source that should be equivalent to the built-in egs\_fano\_source.  Thus, Fano charged particle equilibrium conditions should be established using the new source and application codes.  Particles are generated using both a monoenergetic 500~keV electron source and an egs\_radionuclide\_source $^{177}$Lu electron (all other decay types are filtered out) source.  Fano test simulations are performed for 10$^{10}$ histories to achieve sub-0.1\% uncertainties.

\subsection{DPK Validation of egs\_mird and egs\_radionuclide\_source} \label{sec:Methods-DPK-comparison}

To cross-validate egs\_mird against works in the literature, a comparison of DPK calculations is performed to ensure that egs\_internal\_source and the modified egs\_radionuclide\_source are functioning properly in a TRT regime. A rectilinear 3D Dose Point Kernel (DPK) is generated for three radionuclide sources ($^{131}$I, $^{90}$Y, and $^{177}$Lu) using egs\_mird, the appropriate egs\_radionuclide\_source and an egs\_internal\_source generating source particles isotropically within the center voxel.  Two DPK geometries are used, a 16.5~\cm side length cube with 3~\mm side length cubic voxels (for $^{131}$I and $^{90}$Y) and a 43.6~\cm cube with 4~\mm voxels (for $^{177}$Lu), to compare to MC DPK results found in the literature\cite{Pa09a,Hi15}.  Soft tissue compositions of Cristy and Eckerman\cite{Bo99g} (listed in Table~\ref{tab:mediaComp}) are used throughout the entire cubic volume in both kernels, and egs\_internal\_source combined with an egs\_radionuclide\_source are used to generate particles homogeneously within the center voxel.

\subsection{Full patient simulations} \label{sec:Methods-patients}

Full patient simulations are performed using using inherently registered CT and PET scan data acquired with a PET/CT scanner from a prospective clinical trial (NCT04009174) on men with untreated, biopsy-proven, localized prostate cancer.  The study protocol was approved by the institutional Research Ethics Board and all participants provided written informed consent before the study.  One of the goals of this clinical study was to determine if PET scanning with [$^{18}$F]-DCFPyL can accurately identify cancer nodule(s) within the prostate gland as compared with histopathology of biopsy or the explanted prostate gland.  The [$^{18}$F]-DCFPyL PET scan was done in the dynamic mode where a series of images of the pelvic region (including the prostate) with variable image durations of 10~s (10~images), 20~s (5), 40~s (4), 60~s (4) and 180~s (4) were acquired over 22~minutes. Before the dynamic, 16~cm axial length PET scan, a CT scan of the same pelvic region was also acquired to serve as anatomical guidance for the lower spatial resolution PET images; the patient did not move on the patient couch between the two imaging sessions.  Both CT and PET image slices have a thickness of 3.75 and 3.27~\mm respectively, however the CT voxel size was 0.98x0.98x3.75~\mms$^{3}$ while the PET voxel size was 3.90x3.90x3.27~\mms$^{3}$.  A CT structure file containing the contour data for the patient CT was also generated using the MIM Maestro (MIM Software Inc., OH, USA) auto-contouring software.

Using in-house conversion tools (available at https://github.com/Robarts-Lee-Lab/DICOM\_tools), the clinical DICOM files are converted into an EGSnrc format patient geometry (an egsphant file) and an activity file which contains a list of TIA for each voxel.  The patient geometry is obtained by converting the 512x512 HU data for 47 slices from the DICOM format to a respective density using a CT density curve appropriate for the machine.  Media are then assigned using an assignment scheme approach, used previously with EGSnrc egsphants in a brachytherapy context\cite{Mi17}, with the same density thresholds for different media.  Table~\ref{tab:assScheme} lists all the assignment schemes used for different structures in this work, similar to that of Miksys \etal\cite{Mi17}, with an additional structure for the femurs where the soft tissue is split into three distinct marrow media (yellow, yellow-red, red), defined by Schneider \textit{et al}\cite{Sc00b}.  The media P50C50 is a media assigned to voxels that are determined to be approximately 50\% calcification (by mass), to create a middle ground for voxels that are only partially calcification.  Two phantoms are generated from the CT data, one with marrow media defined in the femurs using the contour data and one where the femurs are treated as generic patient regions (\ie assigned media are `Male Soft Tissue' and `Cortical Bone').  All full patient simulations are performed for 10$^{9}$ histories to achieve $\sim$1\% uncertainties in regions of interest.

\begin{table}[htbp]
  \vspace{0.8\normalbaselineskip}
  \centering
  \captionl{Assignment Schemes}{The media assignment scheme used to assign the media described in Table~\ref{tab:mediaComp} to different structures defined for the patient CT.
  \label{tab:assScheme}}
  \\
    \begin{tabular}{crcc}
    \hline \hline \vspace{-0.8\normalbaselineskip} \\
    \multirow{2}[4]{*}{DICOM Structure} & \multicolumn{1}{c}{\multirow{2}[4]{*}{Medium}} & \multicolumn{2}{c}{Density (\dens)} \\
	\cmidrule{3-4}          &       & Lower & Upper \\
    \midrule
    \multicolumn{1}{c}{\multirow{3}[2]{*}{Patient (everywhere)}} & Air   & 0     & 0.75 \\
          & Male Soft Tissue & 0.75  & 1.14 \\
          & Cortical Bone & 1.14  & $\infty$ \vspace{0.3\normalbaselineskip} \\
    \multirow{3}[2]{*}{Prostate} & Prostate & 0     & 1.14 \\
          & P50C50 & 1.14  & 1.27 \\
          & Calcification & 1.27  & $\infty$ \vspace{0.3\normalbaselineskip} \\
    Rectum & Rectum & 0     & $\infty$ \vspace{0.3\normalbaselineskip} \\
    Bladder & Bladder (full) & 0     & $\infty$ \vspace{0.3\normalbaselineskip} \\
    \multirow{2}[2]{*}{Femur} & Male Soft Tissue & 0  & 1.14 \\
          & Cortical Bone & 1.14  & $\infty$ \\
    \hline \hline \vspace{-0.8\normalbaselineskip} \\
    \end{tabular}
  \vspace{0.8\normalbaselineskip}
\end{table}%

\begin{figure}[htbp]
	\centering
	\includegraphics[width=1.00\textwidth]{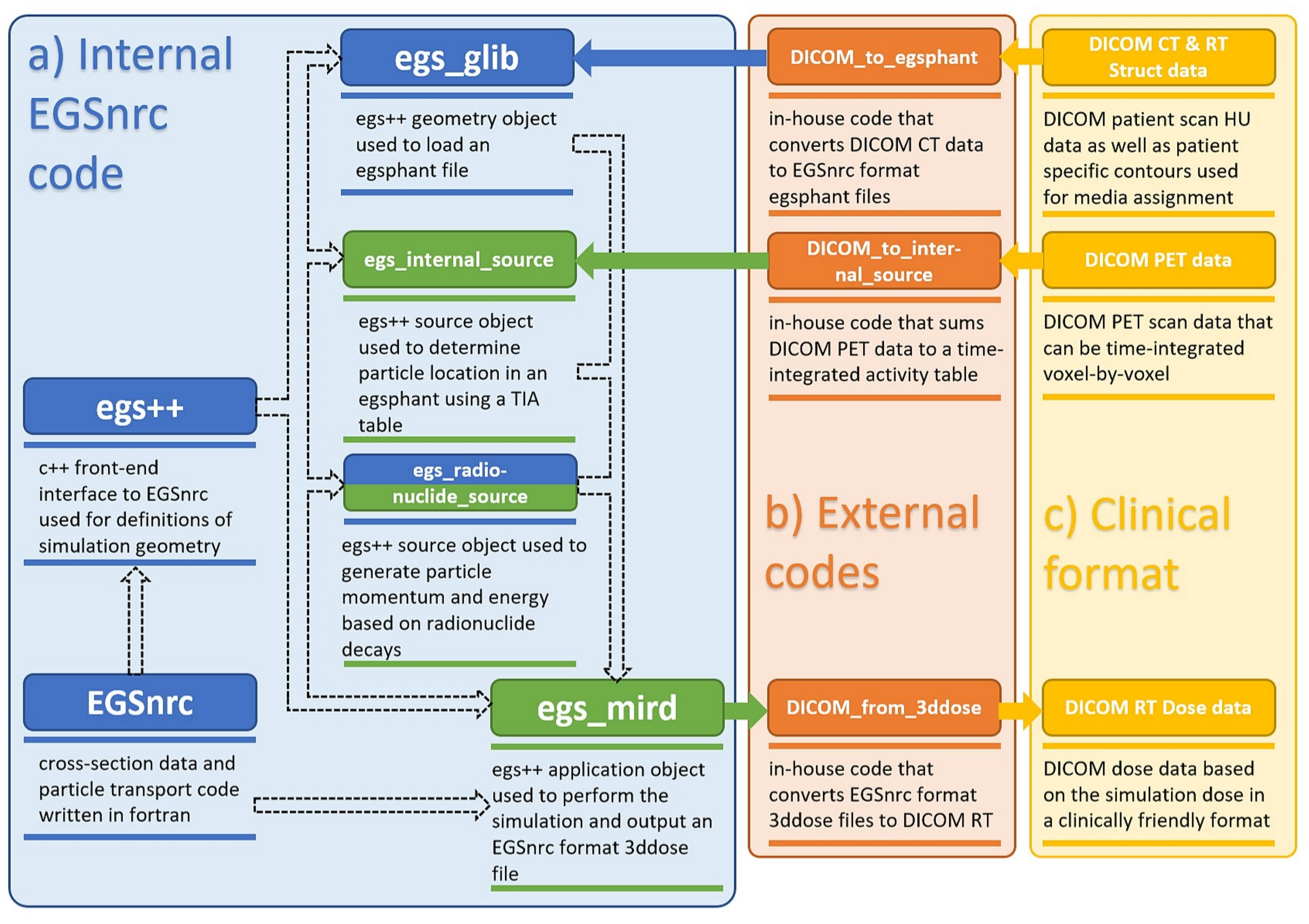}
	\captionl{Workflow}{A flow chart showing the overall structure and workflow of egs\_mird.  Going from clinical DICOM CT and PET data, to an EGSnrc egs\_mird simulation, and back to a DICOM format RT Dose file.  The blue area (a) represent the EGSnrc (blue) and new egs\_mird (green) elements used to perform simulations, the yellow area (c) represents the DICOM data used as input and output, the orange area (b) elements are in-house conversion tools (available at https://github.com/Robarts-Lee-Lab/DICOM\_tools) to go between DICOM and EGSnrc formats.  Solid arrows indicate input/output and dashed arrows indicate code dependency.  All blue area (a) definitions and input/output are handled in a single egsinp (input) file to be created by a user.
	\label{fig:Workflow}}
\end{figure}

The PET data registered to the CT phantom are summed over all acquisition windows (22 minutes total), weighted by the duration of each acquisition and multiplied by voxel volume.  Thus, the sequence of PET scans detailing activity of [$^{18}$F]-DCFPyL in Bq/mL is converted into a single 3D array of TIA.  Due to the resolution of the PET scanner, the 3D TIA array is far coarser than the patient CT data; the CT data has a resolution of 512x512 pixels per slice and the PET data have a resolution of 128x128 pixels per slice.  Thus, to create a list of activity weights across the egsphant based on the CT data, the activities at the center of each egsphant voxel were linearly interpolated using the 3D TIA data.  Additionally, if the CT voxel has a density below 0.75 (assumed to be air), the activity is set to zero.  To investigate the potential effect of calcifications, which can influence dose results substantially in brachytherapy\cite{Mi17,Vi19}, two sets of activities are then generated, one with no further restrictions and another where the activity in calcifications is set to zero (halving the activity in the hybrid prostate/calcification voxels).  Figure~\ref{fig:Workflow} shows an overview of the entire workflow in generating a dose distribution using clinical PET and CT data.

\begin{figure}[htbp]
	\centering
	\includegraphics[width=0.95\textwidth]{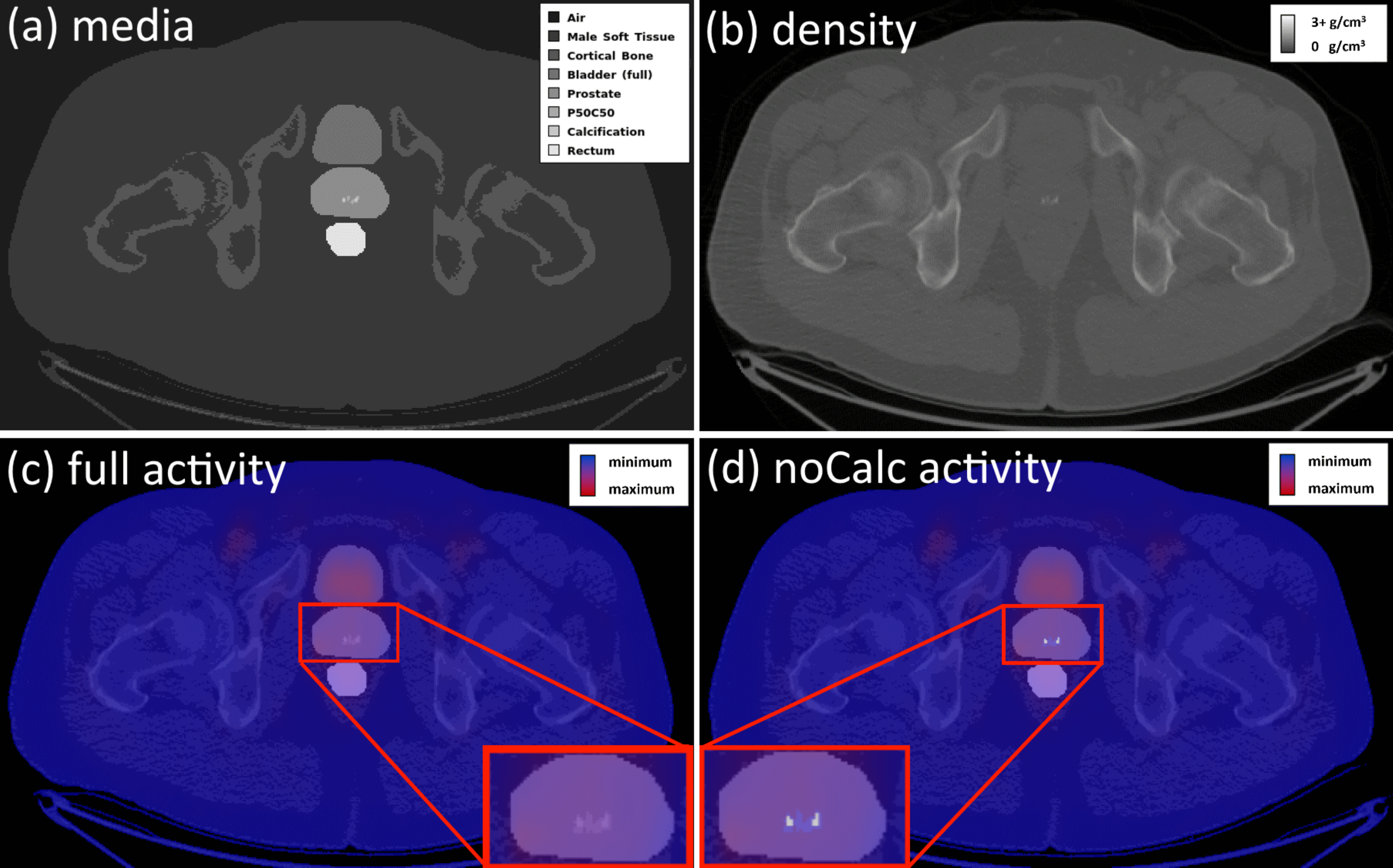}
	\captionl{Four phantom images}{Slice cross-sections showing egsphant (a) media assignment and (b) density assignment, as well as an TIA heat map for (c) full activity and (d) no activity in calcifications based on [$^{18}$F]-DCFPyL PET scan data.  Lowest TIA is in blue and highest TIA is in red, regions with no color have no TIA.
	\label{fig:Phantoms}}
\end{figure}

Two sets of simulations are performed with the egsphants and TIA weight files using a $^{177}$Lu source; one simulation with no variance reduction techniques (noVR) where both photon and electron transport is performed, one simulation where electron transport is not performed (electron energy is deposited locally) and dose due to photon emissions is calculated using TL scoring.  Thus, this work is simulating a $^{177}$Lu activity distribution based on [$^{18}$F]-DCFPyL PET scan data. These simulations are performed using both egsphant models (with and without explicit bone marrow media) as well as the TIA arrays in which the calcifications do and do not have activity.  Figure~\ref{fig:Phantoms} shows a cross-section of the patient that shows the rectum, prostate (with calcifications), and bladder.  The density as well as the full and no calcification TIA maps are also shown.

To compare to the MC patient simulations, an (80~\cms)$^3$ DPK kernel with voxel resolution matching the CT data of the patient (0.0976x0.0976x0.351~\cms$^3$) is generated in prostate tissue (Table~\ref{tab:mediaComp}), similarly to the DPKs described in Section~\ref{sec:Methods-DPK-comparison}. DPK calculations are performed using 3x10$^9$ histories to achieve uncertainties below 0.1\%. The DPK is then convolved with the patient TIA array using MatLab's fast Fourier transform technique to provide a 3D distribution of absorbed dose using the full and no calcification activities.

\section{Results} \label{sec:Results}

\subsection{Validation} \label{sec:Results-validation}
\begin{figure}[htbp]
	\centering
	\includegraphics[width=0.95\textwidth]{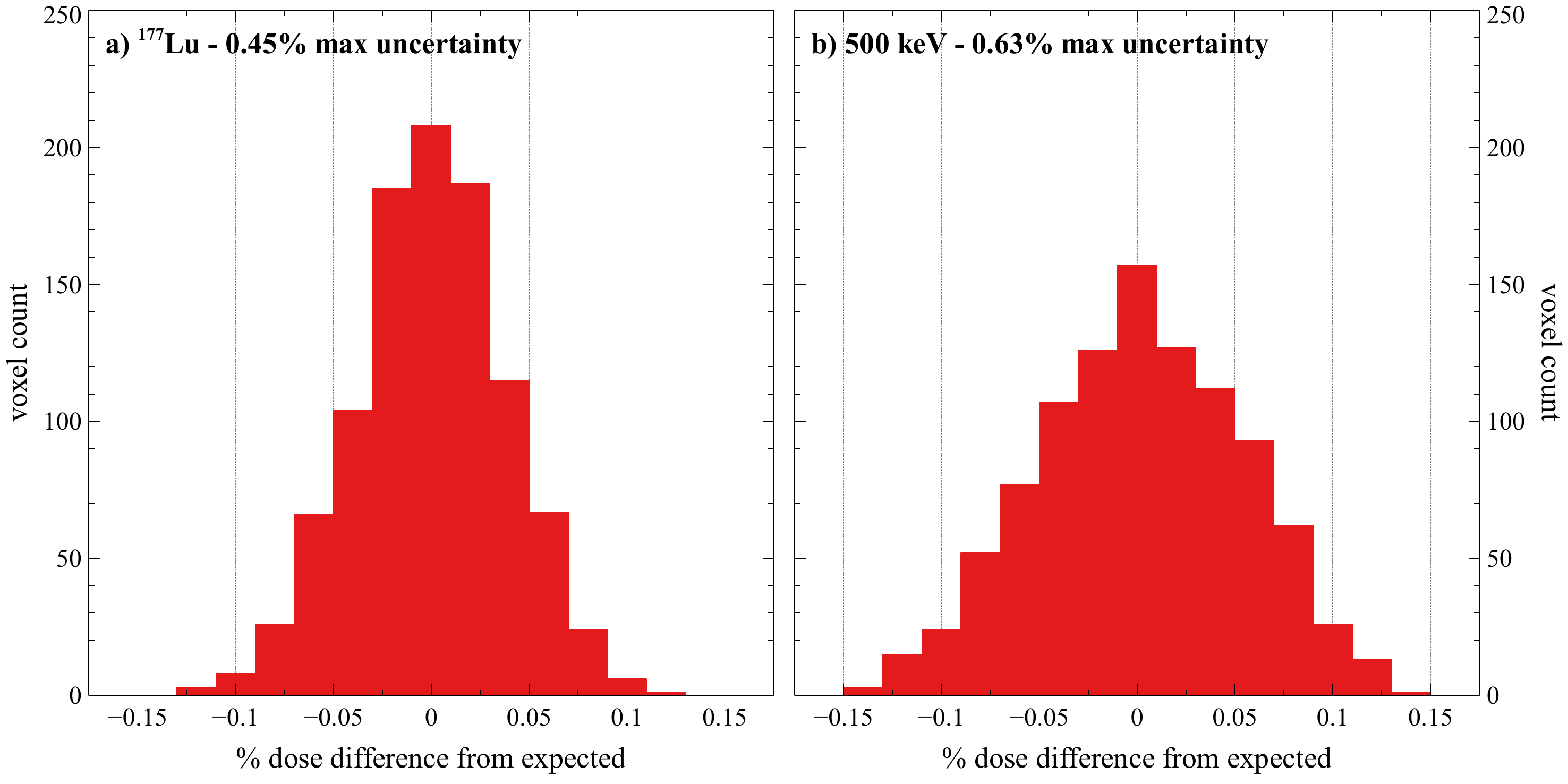}
	\captionl{Fano}{Histogram showing voxel dose difference from the expected Fano test dose for a $^{177}$Lu source (left) and a 500~keV monoenergetic electron source (right).
	\label{fig:Fano}}
\end{figure}

Figure~\ref{fig:Fano} shows two histograms of the percent difference of the 1000 voxel doses and expected Fano dose.  Calculated dose uncertainty is at most 0.063\% for the monoenergetic electrons and 0.045\% for the $^{177}$Lu electrons.  Looking at the overall data, 94\% of the monoenergetic electron doses and 99\% of the $^{177}$Lu electrons are within 0.1\% threshold\cite{Ka99b,Se02}.  Thus, we consider these simulations using egs\_mird, egs\_internal\_source, and egs\_radionuclide\_source as passing the Fano test.

\begin{figure}[htbp]
	\centering
	\includegraphics[width=0.475\textwidth]{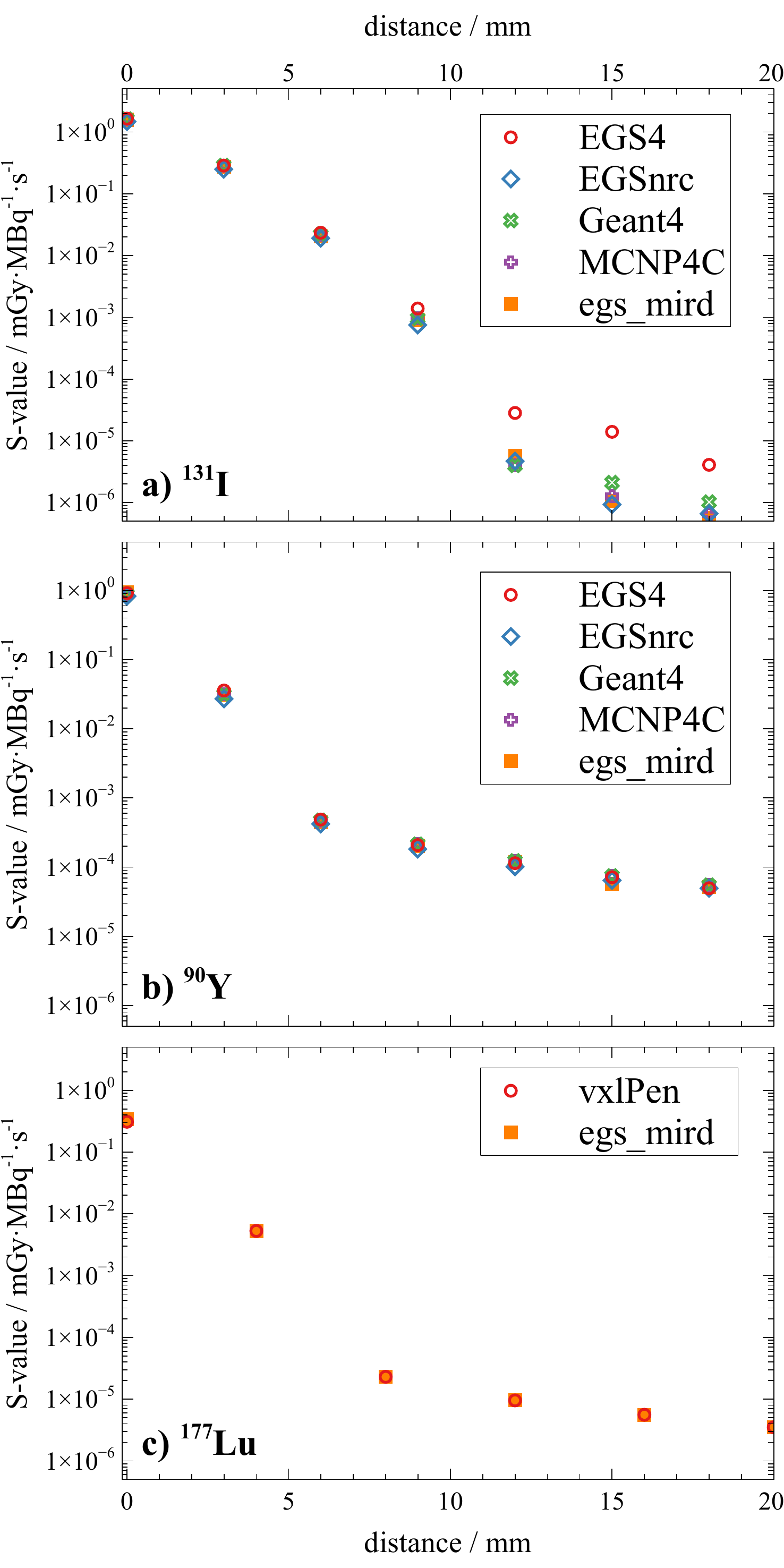}
	\captionl{DPK-Comparison}{Comparison of DPKs generated using egs\_mird (orange) to other DPKs from the literature for $^{131}$I (top)\cite{Pa09a}, $^{90}$Y (middle)\cite{Pa09a}, and $^{177}$Lu (bottom)\cite{Hi15}.  Uncertainties on egs\_mird results are sub-1\%.  Uncertainties in the literature data are either too small to display, or unreported.
	\label{fig:DPK-Comparison}}
\end{figure}

Figure~\ref{fig:DPK-Comparison} shows three scatter plots of DPKs made with three different radioisotopes and compared to other MC results in the literature.  The results computed using the modified egs\_radionuclide\_source, egs\_internal\_source, and egs\_mird align with other results from the literature.  The only exception is the EGS4 results in Figure~\ref{fig:DPK-Comparison}a ($^{131}$I), which deviates at a distance 9~mm or greater.  This deviation is exclusive to EGS4, as it does not occur between egs\_mird and the EGSnrc, Geant4, or MCNP4C results.

\subsection{Patient simulations} \label{sec:Results-patients}
The patient shown in Figure~\ref{fig:Phantoms} is used for the comparisons herein, chosen for the patient's large calcifications.  A tabulated list of computation times is found in Table~\ref{tab:Timing}, for noVR and TL simulations.  All of the in-house developed DICOM conversion code finish in under one minute, collectively.  The simulations themselves take the longest time, 13-15 minutes on the higher end, with file Input/Output (I/O) taking a large fraction of the time.  Simulations are performed on 22 cores of a Ryzen 9 3900XT.  The 22 separate simulation I/O calls push the read/write ability of the solid state drive to its limit, causing a variable I/O bottleneck in the total time.

\begin{table}[htbp]
  \vspace{0.8\normalbaselineskip}
  \centering
  \captionl{Timing}{The time required to convert CT and PET DICOM data to an EGSnrc friendly format, to perform the $^{177}$Lu simulations (split into simulation runtime and input/output time), and to convert the egs\_mird 3ddose file output back into a DICOM format.
  \label{tab:Timing}}
    \begin{tabular}{lcc}
    \hline \hline \vspace{-0.8\normalbaselineskip} \\
    \multirow{2}[4]{*}{process} & \multicolumn{2}{c}{time (min)} \\
\cmidrule{2-3}          & noVR & TL \\
    \midrule
    DICOM CT conv. & \multicolumn{2}{c}{0.51} \\
    DICOM PET conv. & \multicolumn{2}{c}{0.36} \\
    \midrule
    Simulation I/O* & \multicolumn{2}{c}{2-4} \\
    Simulation runtime\textsuperscript{\textdagger} & 28.47 & 11.50 \\
    \midrule
    DICOM RTDOSE conv. & \multicolumn{2}{c}{0.19} \\
    \hline \hline \vspace{-0.8\normalbaselineskip} \\
    \end{tabular} \\
	{\parbox{16.5cm}{
	  * Input and output time is highly variable, averaging roughly 3 minutes \\
	  \textsuperscript{\textdagger} Simulations performed on a single computer, using 22 cores of a Ryzen 9 3900XT}}
  \vspace{0.8\normalbaselineskip}
\end{table}%

\begin{figure}[htbp]
	\centering
	\includegraphics[width=0.95\textwidth]{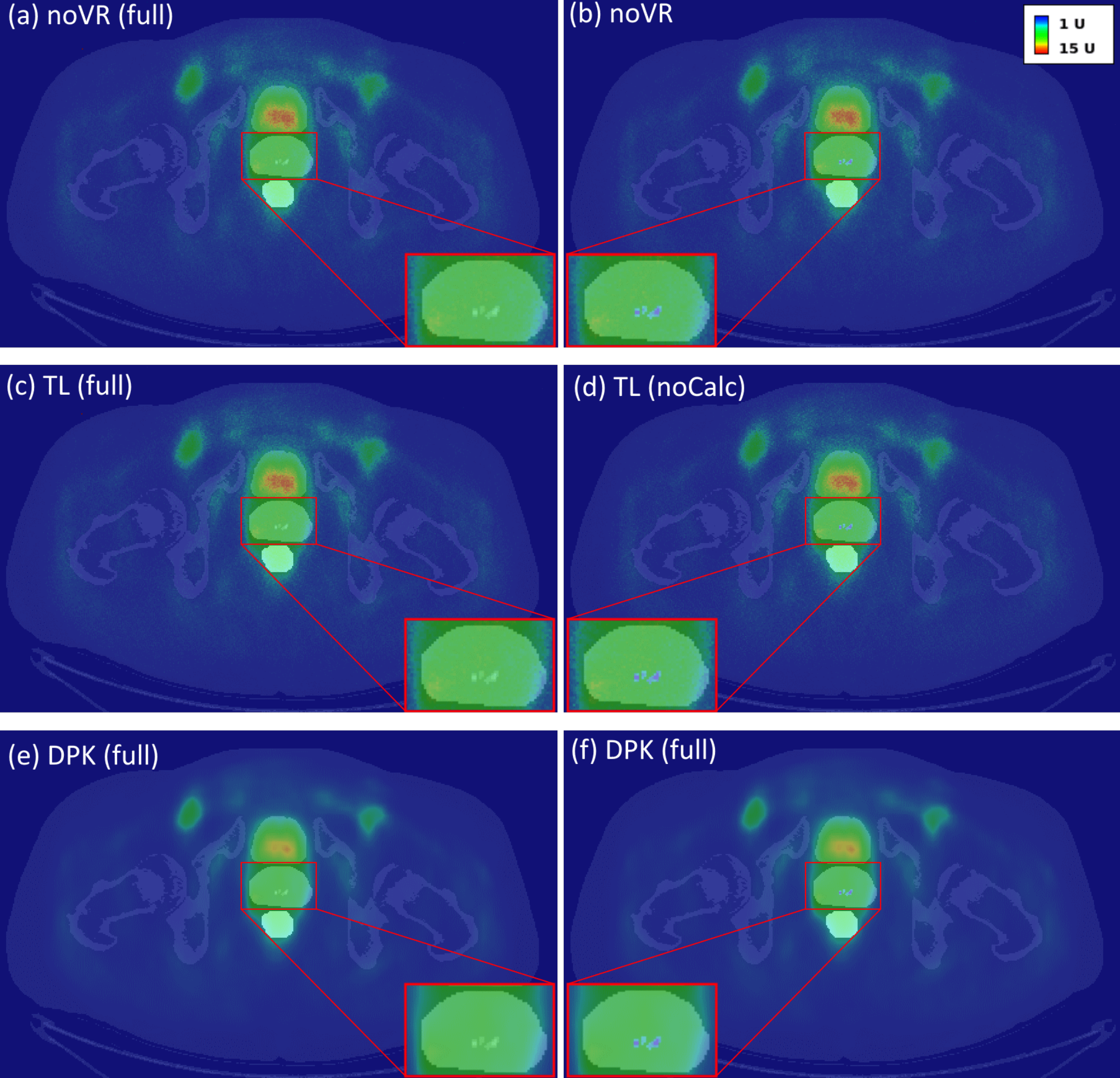}
	\captionl{Six color maps}{Slice cross-sections showing a dose color map in U~=~\uGy of a simulated $^{177}$Lu TRT patient.  Results for the full patient time-integrated activity (left) were calculated using a noVR egs\_mird simulation (top), an egs\_mird simulation using track-length scoring (middle), and a convolution calculation using a kernel calculated with egs\_mird (bottom).  Results for the time-integrated activity with no source in the calcifications are also displayed (right).  A blown-up image of the prostate for each case is also displayed.
	\label{fig:Washes}}
\end{figure}

Figure~\ref{fig:Washes} depicts a dose color map for three different dose calculation methods normalized from Gy/history to \uGy of the patient data in Figure~\ref{fig:Phantoms} for two TIA matrices; one where all PET activity is used (full) and one where the activity in voxels designated as calcification is set to zero and the activity in voxels designated as half-calcification/half-prostate is reduced to half (noCalc).  The noVR color maps (a,b) show the results for the simulations where both electron and photon transport are performed with no variance reduction techniques.  The TL color maps (c,d) show the results for the simulations where photon dose was scored using track-length scoring and all electron energy is deposited in the voxel in which the electron is generated (\ie no electron transport).  The color maps of the two simulations look very similar.  The dose to prostate tissue between color maps with and without calcification TIA look quite similar in all non-calcification regions with an expected drop in dose to the calcifications themselves.  Figure~\ref{fig:Washes}.e and \ref{fig:Washes}.f show the dose results when convolving an egs\_mird calculated $^{177}$Lu DPK, such as the one in Section~\ref{sec:Methods-validation}, with the TIA.  The DPK color maps are much smoother, with lower overall uncertainties.

\begin{figure}[htbp]
	\centering
	\includegraphics[width=0.95\textwidth]{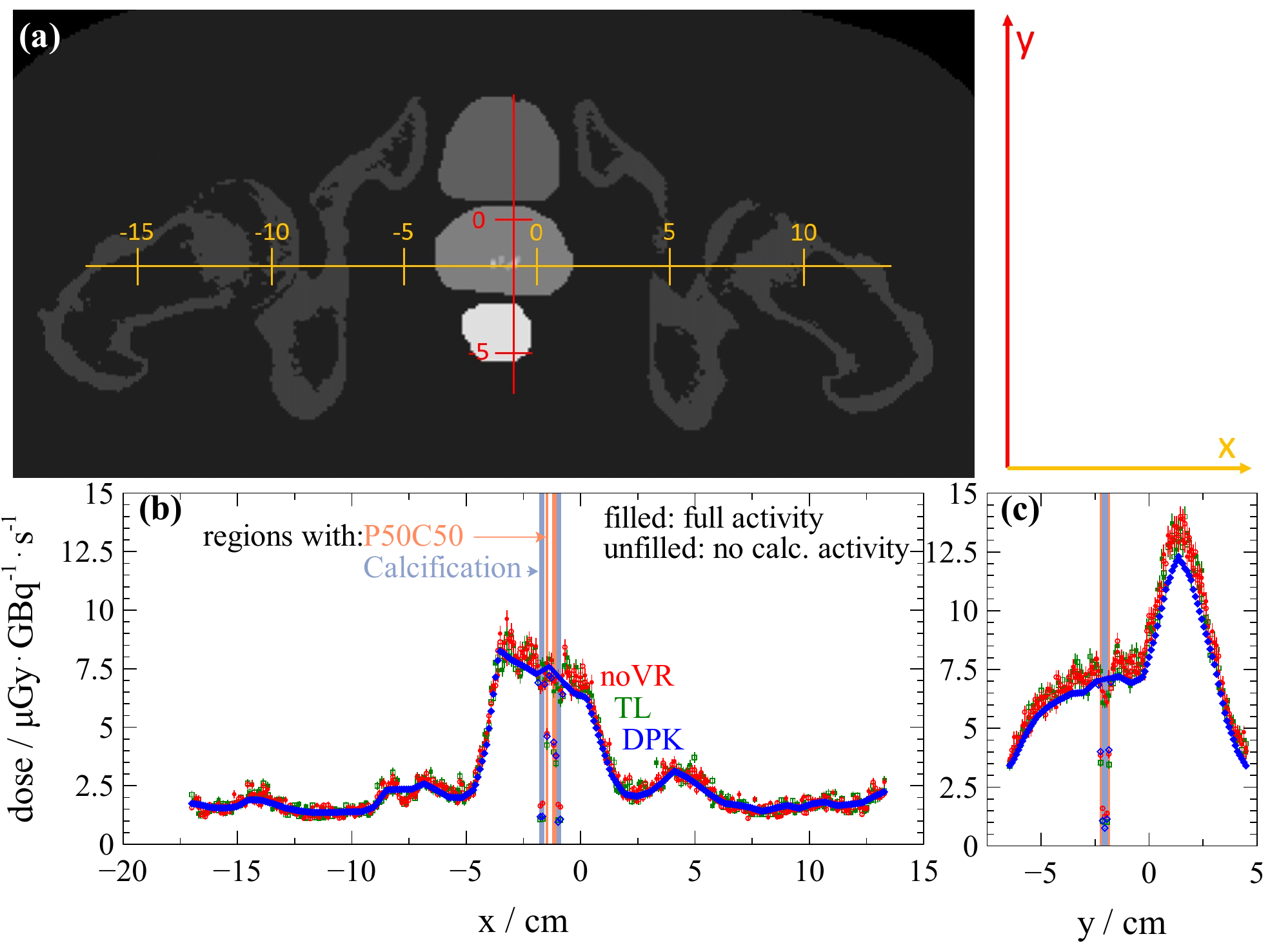}
	\captionl{Profile}{Dose profiles along the x (b) and y (c) axes at the positions shown in cross-section (a) normalized to \uGy.  Results for $^{177}$Lu noVR MC (red), TL MC (green), and DPK convolution (blue) are all plotted.  Markers which are filled represent dose when using total TIA and unfilled markers represent dose when TIA in the calcifications is not simulated; additionally, voxels that contain P50C50 and calcification are shaded light orange and blue, respectively.  All displayed uncertainties are derived from MC dose scoring uncertainty.  The fractional uncertainty of the central voxel dose in the DPK is used for all error bars in the plot.
	\label{fig:Profiles}}
\end{figure}

Two dose profiles going through the prostate and several calcifications, as well as neighboring regions, are plotted in Figure~\ref{fig:Profiles}.  Comparing the four different egs\_mird computations, noVR and TL simulation for both full or noCalc activity, results agree within the uncertainty of the computed MC dose for all non-calcification regions.  There is a local decrease in dose in the calcifications for the noVR and TL simulations, but a much larger decrease when activity in calcifications is removed; a drop of over 80\% in the full calcifications and a drop of almost 40\% in the half-calcification/half-prostate regions.  Within calcification regions, the noVR and TL simulations do not agree within uncertainty, with a significantly lower dose in the TL simulations.  As in the color maps, the DPK profiles appear much smoother.  Additionally, the DPKs are consistently underdosing compared to MC in the prostate region of Figure~\ref{fig:Profiles}a (-3.5 to 0.5~\cms) and the bladder region of Figure~\ref{fig:Profiles}b (0.5 to 2.5~\cms).  In the calcification regions, the TL and DPK results agree far more closely with each other than with the noVR results.

\begin{figure}[htbp]
	\centering
	\includegraphics[width=0.95\textwidth]{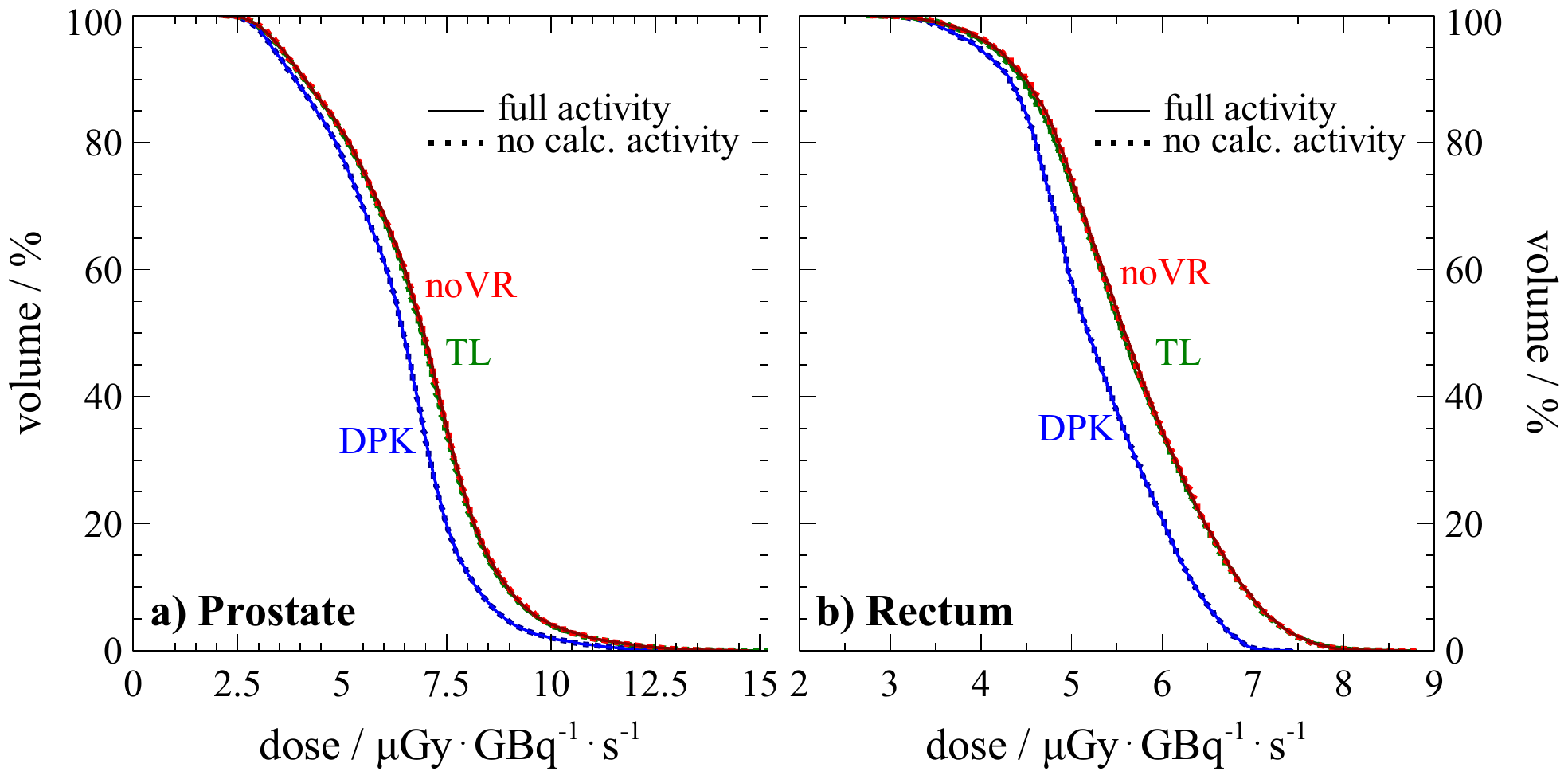}
	\captionl{DVH}{Dose volume histograms showing \uGy for the $^{177}$Lu noVR MC (red), TL MC (green), and DPK convolution (blue) calculations using TIA matrices with full (solid) and noCalc (dashed) activity.  Results for the same calculation with and without activity in the calcifications lie directly atop one another.
	\label{fig:DVHs}}
\end{figure}

Figure~\ref{fig:DVHs} shows DVHs for all calculation methods in all prostate tissue and all rectal tissue (a nearby organ-at-risk).  In all MC simulation cases, DVHs are in rough agreement with very little deviation.  Table~\ref{tab:Dx} lists many dose metrics for the prostate and rectum, further demonstrating agreement between the four different MC simulations.  When including the DPK results, Figure~\ref{fig:DVHs} demonstrates the underdosing seen in Figure~\ref{fig:Profiles} more clearly.  The DPK dose is 7.09\% lower than the noVR dose in the prostate and 6.55\% lower than noVR in the rectum.

\begin{table}[htbp]
  \vspace{0.8\normalbaselineskip}
	\centering
	\captionl{Dose Metrics}{Dx (\uGys{}) and Vx (\% of total volume) metrics in the prostate (with and without including calcifications) and rectum.  Results are shown for the scenario where prostate calcifications have the full activity assigned from the PET data conversion.  Metrics where calcification activity is lowered do not vary significantly.
	\label{tab:Dx}}
	\footnotesize
	\begin{tabular}{ccllllll}
    \hline \hline \vspace{-0.7\normalbaselineskip} \\
    \multirow{2}[4]{*}{Region} & \multirow{2}[4]{*}{Model} & \multicolumn{6}{c}{Metrics} \\
	\cmidrule{3-8} & & \multicolumn{1}{c}{D10} & \multicolumn{1}{c}{D50} & \multicolumn{1}{c}{D90} & \multicolumn{1}{c}{V1} & \multicolumn{1}{c}{V5} & \multicolumn{1}{c}{V10} \\
    \midrule
    \multicolumn{1}{c}{\multirow{3}[2]{*}{Prostate}}                  & Full  & 8.96 & 6.94 & 4.10 & 55.24 & 45.19 & 2.27 \\
                                                                      & TL    & 8.92 & 6.91 & 4.08 & 55.24 & 45.07 & 2.20 \\
                                                                      & DPK   & 8.22 & 6.46 & 3.87 & 55.24 & 43.02 & 1.10 \\
    \midrule
    \multicolumn{1}{c}{\multirow{3}[2]{*}{Prostate \& Calcification}} & Full  & 8.95 & 6.94 & 4.10 & 55.44 & 45.38 & 2.27 \\
                                                                      & TL    & 8.92 & 6.92 & 4.08 & 55.44 & 45.27 & 2.20 \\
                                                                      & DPK   & 8.21 & 6.47 & 3.88 & 55.44 & 43.21 & 1.10 \\
    \midrule
    \multicolumn{1}{c}{\multirow{3}[2]{*}{Rectum}}                    & Full  & 6.89 & 5.59 & 4.49 & 29.33 & 21.89 & 0.00 \\
                                                                      & TL    & 6.86 & 5.56 & 4.47 & 29.33 & 21.68 & 0.00 \\
                                                                      & DPK   & 6.37 & 5.18 & 4.32 & 29.33 & 17.01 & 0.00 \\
    \hline \hline \vspace{-0.7\normalbaselineskip} \\
    \end{tabular}
  \vspace{0.8\normalbaselineskip}
\end{table}

\section{Discussion} \label{sec:Discussion}

The novel EGSnrc application egs\_mird can perform a full patient TRT treatment simulation using raw DICOM data within 15 minutes, producing an RTDOSE file with resolution matching a 512x512x47 voxel CT phantom.  Downsampling the CT resolution to match the PET resolution in this scenario could offer further speed gains, pushing the whole process to be completed within 10 minutes.  There are no significant differences between a full patient simulation with electron and photon transport and a simulation with no electron transport and using track-length scoring for photon dose.  Both types of simulations provide similar doses and uncertainties in areas with high TIA, but TL simulations have lower uncertainties in regions where photons were the primary dose contributors in addition to also finishing twice as fast when not modelling electron transport.

As shown in Section~\ref{sec:Results-validation}, when performing S-value kernel calculations using egs\_internal\_source and egs\_radionuclide\_source for $^{131}$I, $^{90}$Y, and $^{177}$Lu radionuclide sources, egs\_mird results in Figure~\ref{fig:DPK-Comparison} are comparable to the literature.  Additionally, simulations replicating egs\_fano\_source using egs\_internal\_source in Figure~\ref{fig:Fano} demonstrates the reliability of both egs\_internal\_source and egs\_mird, in the context of centimeter-sized voxels.

\begin{figure}[htbp]
	\centering
	\includegraphics[width=0.95\textwidth]{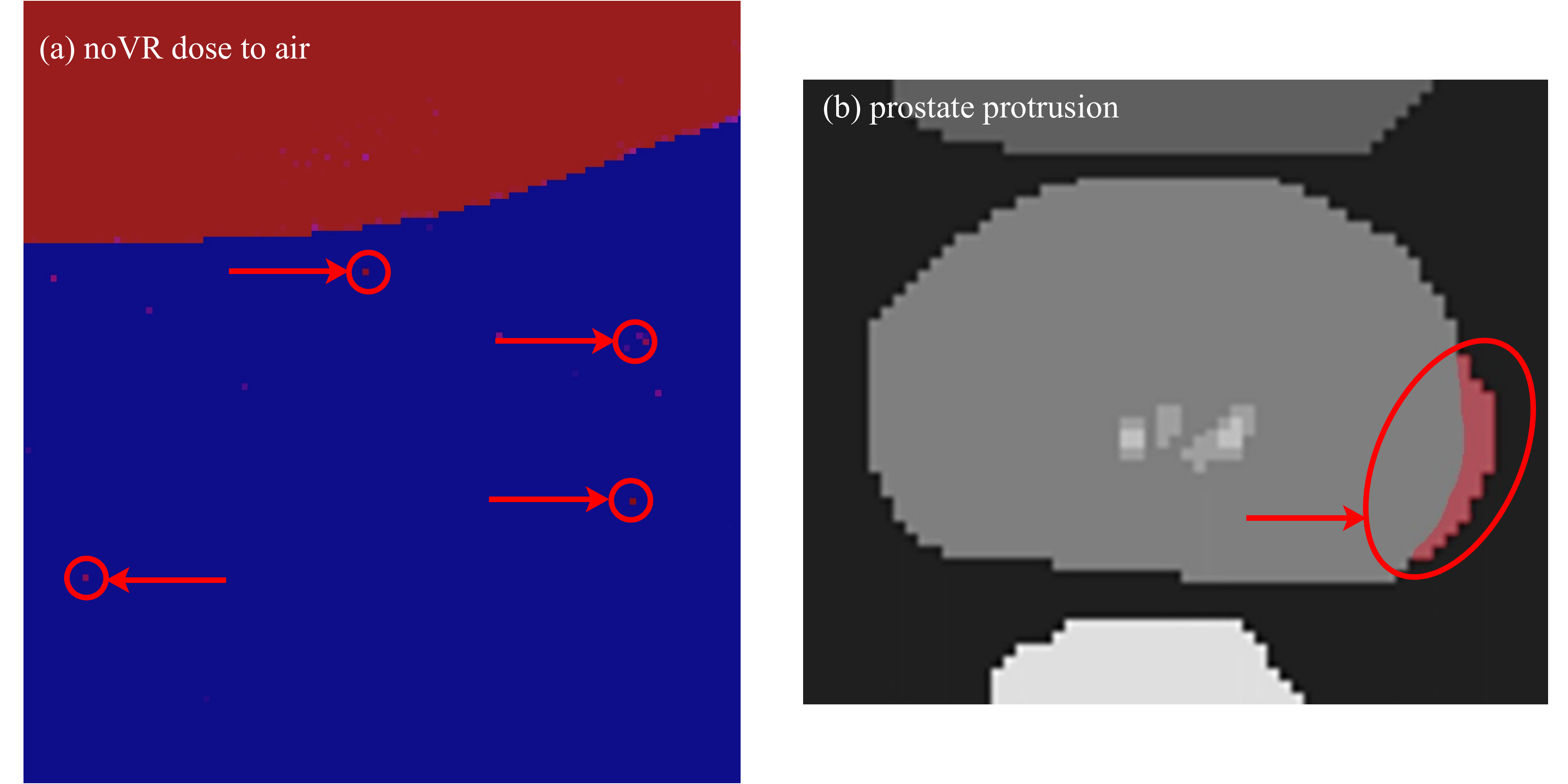}
	\captionl{Zoom}{Zoomed in crops of Figure~\ref{fig:Washes} (a) showing high dose air voxels outside of the patient for noVR MC results of $^{177}$Lu simulations and Figure~\ref{fig:Phantoms} (b) showing the prostate protrusion to the patient's bottom left (image's bottom right) of the prostate.
	\label{fig:Zooms}}
\end{figure}

All full patient dose calculation results in this work are calculated for $^{177}$Lu TRT treatment.  As this work is meant as an introduction to and validation of egs\_mird, this study's scope is limited to the investigation of a single beta and gamma particle emitter.  The focus of this work is to investigate different MC transport options in egs\_mird, choices in modelling approach when translating patient CT and PET data to an EGSnrc simulation, and comparing egs\_mird results to a DPK convolution method.  Simulations comparing patient dose as a function of TIA for different radionuclides using the same CT and PET data would be an interesting investigation for future study, but is too broad for this introductory work.

A second patient model was used for simulations with a different tissue assignment in the femur than that indicated in Table~\ref{tab:assScheme}.  Femur soft tissue was assigned a mix of three different marrow compositions (listed in Table~\ref{tab:mediaComp}) within the patient femurs instead of male soft tissue.  There was no significant difference found in dose to the soft tissue volume in the femurs, whether using bone marrow compositions or generic male soft tissue.  Thus, as bone marrow is not a primary focus of this study, and it lies in a region with low TIA, male soft tissue is used throughout this work for simplicity.  In Figure~\ref{fig:Zooms}b, focusing on the prostate contour region, there are observations to note.  On careful examination, one can see that the prostate contour used for media assignment has a protrusion in the expected shape on the patient's lower left (image's lower right) side.  This region corresponds to low TIA in the PET data, which also corresponds to a low dose region apparent in Figure~\ref{fig:Washes}, where in all cases, one can see a clear lower dose region (below 8~\uGys{}) on the patient's bottom-left side.  Though likely due to imprecise contouring, it is possible that there could also have been image registration issues or patient anatomy or position changes over the full acquisition time.  If this region is indeed misassigned prostate, then the low doses in the DVH in Figure~\ref{fig:DVHs}a are not in the actual prostate volume, and one would expect a higher minimum dose to 100\% of the properly assigned prostate volume.

In further comparison of the CT and PET data, it is apparent that the calcifications which are very evident in the CT scan, do not appear in the PET data; this is due, in part, to the low spatial resolution of the PET scan relative to the CT scan.  Beyond the lower spatial resolution, the final 3D TIA matrix is also an averaged TIA over a 22 minute acquisition, if the patient's prostate exhibits shifts on the order of a few \mm (the size of the calcifications), then a lack of TIA could be time-averaged out when computing the total TIA in each voxel.

Figure~\ref{fig:Washes} shows a color map comparison of the four different $^{177}$Lu Monte Carlo simulations, showing little difference between the maps of the noVR simulations and the TL simulations.  This is further supported with the agreement between noVR and TL simulations demonstrated in Figures~\ref{fig:Profiles} and \ref{fig:DVHs}.  Thus, when performing prostate TRT simulations, electron transport is not necessary.  This follows from the logic that there is TIA throughout the patient with very shallow TIA and density gradients between most voxels, setting up near charged particle equilibrium conditions throughout most of the phantom.  The only notable difference between the distributions is the high dose areas in the air outside the patient in the full simulation (compared to TL), sparsely seen in Figure~\ref{fig:Washes}a~and~b, but outlined more clearly in Figure~\ref{fig:Zooms}a.

These very high dose (and associated high uncertainty) voxels in the air are caused by the small number of photons escaping the patient in the simulation.  When performing interactive photon scoring in these simulations,  those few photons that do escape can often travel far in the low density air before interacting.  Thus, in the case of the noVR simulations, the vast majority of air voxels score zero dose as no photons interacted within them, a small number of voxels score dose from a single interaction event (with 100\% associated uncertainty as a single event does not allow for any statistics), and very few voxels score dose due to multiple interactions (which would have a more reasonable dose estimate and associated uncertainty).  If the number of histories were increased by several orders of magnitude (and simulations took at least 1000 times longer), such that each air voxel around the patient should expect several photon interaction events, most of the air voxels would have a reasonable dose estimate and uncertainty.

Comparing Figure~\ref{fig:Washes}e,f to the MC results, DPK results are markedly smoother; a similar smoothness in the DPK dose profiles (compared to MC) can be observed in Figure~\ref{fig:Profiles}.  Additionally, Figure~\ref{fig:DVHs} also demonstrates a systematically lower dose for DPK results compared to MC.  These discrepancies between the models stem from several factors.  Firstly, the DPK is calculated to very low uncertainties (sub-0.1\%, compared to patient MC uncertainties which are on the order of a few percent.  Secondly, the DPK does not use the CT data, which itself is very heterogeneous and introduces many of the perturbations in the MC results.  Finally, the only large contributing noise the DPK contends with is that of the PET scan, but the PET scan activity itself was up-sampled from 128x128 to 512x512 resolution using linear interpolation, also increasing smoothness between neighbouring voxels.  These factors all combine together to make for a very smooth distribution compared to MC.

Looking at the systemic discrepancies between DPK and MC, there are several contributing factors.  DPK results do not incorporate CT data in any way, assuming all voxels are essentially prostate tissue with a density of 1.04~\dens.  In comparison, the converted CT results in the MC have a prostate, rectum, and bladder density of 1.01, 1.02, and 1.00~\dens, respectively.  Furthermore, even using the large 80~cm sidelength kernel, roughly 1\% of the particles (by energy) generated in the center voxel escape the DPK.  Finally, roughly 22\% of $^{177}$Lu decays from the source are non-beta decays, and for many of the possible photon decays, the change in mass energy absorption coefficients with energy and composition\cite{Mi16} can further increase the discrepancy between the DPK and MC results.

To verify that the $^{177}$Lu DPK and MC calculations agree when removing all the above discrepancies, a MC simulation is performed where no electron or photon transport is performed and all energy is locally deposited.  This simulation is compared to a DPK convolution using a DPK that also includes no transport (all energy is deposited in the center voxel).  The DPK results were then normalized to the voxel mass used in the MC simulation.  In all regions of interest investigated, these two models now agreed within 0.1\%, far smaller than the 7.09\% seen in Figure~\ref{fig:DVHs}.

As the TL distribution matches the noVR simulation and is less noisy in regions with no TIA (regions receiving primarily photon dose), it is recommended that TL scoring option (which runs in roughly half the time) is used when applicable.  Though, as stated above, it is reliable when the TIA and voxel density gradients are not large, it is not recommend that TL scoring is used when using a very heterogeneous TIA distribution or when modelling a phantom with a very heterogeneous media/density distribution.  For example, it is highly recommended that a full simulation with no variance reduction be used for TRT treatments of lung carcinoma, as the very low density in the lung and around the tumour will affect electron transport (and this electron transport should be modelled).

Figure~\ref{fig:Profiles} shows line profiles along the x- and y-axis which go through prostate calcifications.  The fluctuation of dose along the profiles is often larger than the shown uncertainties (which are solely the statistical uncertainties of the MC dose calculations), they do not account for the noise inherent in the image data.  The voxels with pure prostate tissue (even those voxels which neighbor calcifications) received the same dose whether or not prostate calcifications were assigned activity in the simulations.  Thus, even if calcifications have minimal radionuclide uptake, the presence of calcifications will not impact pure prostate tissue voxels.

As for the voxels assigned P50C50 (50/50 mix by mass of prostate and calcification), their dose does vary with the change to activity in the calcification, showing a 40\% drop when full calcification TIA is to zero and P50C50 is halved.  Due to the limited spatial resolution of the patient model, the energy deposition to the prostate fraction and to the calcification fraction of voxel mass cannot be decoupled.  This merits further investigation with higher spatial resolution data.  It is of note that the pure calcification voxels do still have some small dose even when TIA in them is set to zero, indicating that there is at least some dose contribution from surrounding non-calcification voxels.  This is a strong indication that the reduced energy deposited in the P50C50 voxels would not be entirely localized within the prostate fraction of the voxel media if the different media in the voxel were to be modeled explicitly.

As explored in this work, there are many considerations when performing full patient dose calculations in TRT.  Although some model details, such as the need to model bone marrow explicitly, are shown to be unneeded in this case, this study (limited to a single patient prostate treatment) still found some substantial differences between the MC approach and the DPK convolution.  These differences may become even more substantial when investigating TRT treatments in regions with higher tissue heterogeneity (such as the lung).  And, though the simple DPK used herein could calculate results within a few minutes, the full MC simulation only takes 15 minutes and provides much more insight.  Thus it is recommended that any clinical trials or implementations of TRT moving forward use MC calculations (with track-length scoring when possible), for which egs\_mird is an excellent candidate.

\section{Conclusion} \label{sec:Conclusion}

This work has introduced egs\_mird, along with egs\_internal\_source, a fast EGSnrc MC code which can calculate a full patient dose distribution expected in $^{177}$Lu TRT using analogous patient CT (47 512x512 slices covering an axial length of 16~cm) and [$^{18}$F]-DCFPyL PET data within 15 minutes.  Even in the relatively homogeneous case of prostate $^{177}$Lu TRT, egs\_mird results were roughly 7\% higher on average than a DPK approach using the exact same patient data, with much larger discrepancies in areas of the patient where media was markedly different from normal tissue (such as bone and calcifications).  Though more advanced DPK models than that explored in this paper are possible, further complicating a simple model (and potentially slowing it down) might not be worth it when a far more robust code such as egs\_mird approaches a similar time window.  Furthermore, egs\_mird could also be used for far more complicated cases where DPK convolution would simply not be suited (such as tumours close to bone or lung tissue regions), helping both with further development of TRT and potential clinical implementations in the future.

\section*{Acknowledgments}

The authors acknowledge support from the Canada Foundation for Innovation's Innovation Fund, the Ontario Innovation Fund, the Ontario Institute of Cancer Research, the Natural Sciences and Engineering Research Council of Canada (NSERC) [funding reference number 06267-2016], Canada Research Chairs (CRC) program, an Early Researcher Award from the Ministry of Research and Innovation of Ontario, and the Carleton University Research Office, as well as access to computing resources from Compute/Calcul Canada.  The authors declare that they do not have any conflict of interest.

\setlength{\baselineskip}{0.55cm}

\section{Bibliography}
\begin{thebibliography}{10}

\bibitem{Mo20}
E.~Mora-Ramirez et~al.,
\newblock Comparison of commercial dosimetric software platforms in patients
  treated with 177Lu-DOTATATE for peptide receptor radionuclide therapy,
\newblock Med. Phys. {\bf 47}, 4602--4615 (2020).

\bibitem{Ka99a}
I.~Kawrakow,
\newblock { Accurate condensed history Monte Carlo simulation of electron
  transport. I. EGSnrc, the new EGS4 version},
\newblock Med. Phys. {\bf 27}, 485 -- 498 (2000).

\bibitem{Vo20}
J.~Violet et~al.,
\newblock Long-term follow-up and outcomes of retreatment in an expanded
  50-patient single-center phase II prospective trial of 177Lu-PSMA-617
  theranostics in metastatic castration-resistant prostate cancer,
\newblock Journal of Nuclear Medicine {\bf 61}, 857--865 (2020).

\bibitem{Ho21}
M.~S. Hofman et~al.,
\newblock [177Lu] Lu-PSMA-617 versus cabazitaxel in patients with metastatic
  castration-resistant prostate cancer (TheraP): a randomised, open-label,
  phase 2 trial,
\newblock The Lancet {\bf 397}, 797--804 (2021).

\bibitem{Vo19}
J.~Violet et~al.,
\newblock Dosimetry of 177Lu-PSMA-617 in metastatic castration-resistant
  prostate cancer: correlations between pretherapeutic imaging and whole-body
  tumor dosimetry with treatment outcomes,
\newblock Journal of Nuclear Medicine {\bf 60}, 517--523 (2019).

\bibitem{Sa14}
{D. Sarrut {\em et al}},
\newblock {A review of the use and potential of the GATE Monte Carlo simulation
  code for radiation therapy and dosimetry applications},
\newblock Med. Phys. {\bf 41}, 064301 (14pp) (2014).

\bibitem{Sa17}
D.~Sarrut, J.-N. Badel, A.~Halty, G.~Garin, D.~Perol, P.~Cassier, J.-Y. Blay,
  D.~Kryza, and A.-L. Giraudet,
\newblock 3D absorbed dose distribution estimated by Monte Carlo simulation in
  radionuclide therapy with a monoclonal antibody targeting synovial sarcoma,
\newblock EJNMMI physics {\bf 4}, 1--16 (2017).

\bibitem{Bi14}
R.~P.~Baum,
\newblock Therapeutic nuclear medicine
\newblock Springer, 2016.

\bibitem{Ka05a}
I.~Kawrakow,
\newblock {egspp: the EGSnrc C++ class library},
\newblock Technical Report PIRS--899, National Research Council Canada, Ottawa,
  Canada, 2005.

\bibitem{Mc16}
A.~McNamara, W.~Kam, N.~Scales, S.~McMahon, J.~Bennett, H.~Byrne, J.~Schuemann,
  H.~Paganetti, R.~Banati, and Z.~Kuncic,
\newblock Dose enhancement effects to the nucleus and mitochondria from gold
  nanoparticles in the cytosol,
\newblock Phys. Med. Biol. {\bf 61}, 5993--6010 (2016).

\bibitem{To18}
R.~Townson, F.~Tessier, and R.~Galea,
\newblock EGSnrc calculation of activity calibration factors for the Vinten
  ionization chamber,
\newblock Applied Radiation and Isotopes {\bf 134}, 100--104 (2018).

\bibitem{ICRU46}
ICRU,
\newblock {Photon, Electron, Photon and Neutron Interaction Data for Body
  Tissues},
\newblock ICRU Report~46, ICRU, Washington D.C., 1992.

\bibitem{Sc00b}
W.~Schneider, T.~Bortfeld, and W.~Schlegel,
\newblock Influence of tissue composition on the results of Monte Carlo
  simulations for patient dose calculations,
\newblock in {\em The Use of Computers in Radiation Therapy}, pages 443--445,
  Springer, 2000.

\bibitem{Ri04}
M.~J. Rivard, B.~M. Coursey, L.~A. DeWerd, M.~S. Huq, G.~S. Ibbott, M.~G.
  Mitch, R.~Nath, and J.~F. Williamson,
\newblock { Update of AAPM Task Group No. 43 Report: A revised AAPM protocol
  for brachytherapy dose calculations},
\newblock Med. Phys. {\bf 31}, 633 -- 674 (2004).

\bibitem{Cr87}
M.~Cristy and K.~Eckerman,
\newblock Specific Absorbed fractions of energy at various ages from internal
  photon sources: 1, Methods,
\newblock Technical report, Oak Ridge National Lab., TN (USA), 1987.

\bibitem{Wi87b}
J.~F. Williamson,
\newblock { Monte Carlo evaluation of kerma at a point for photon transport
  problems},
\newblock Med. Phys. {\bf 14}, 567 -- 576 (1987).

\bibitem{Ch16}
M.~Chamberland, R.~Taylor, D.~Rogers, and R.~Thomson,
\newblock egs\_brachy: a versatile and fast Monte Carlo code for brachytherapy,
\newblock Phys. Med. Biol. {\bf 61}, 8214 (2016).

\bibitem{KR79a}
R.~A. Kronmal and A.~V. Peterson~Jr,
\newblock On the alias method for generating random variables from a discrete
  distribution,
\newblock The American Statistician {\bf 33}, 214--218 (1979).

\bibitem{MT20}
M.~P. Martinov and R.~M. Thomson,
\newblock Technical Note: Taking EGSnrc to new lows: Development of egs++
  lattice geometry and testing with microscopic geometries,
\newblock Med. Phys. {\bf 47}, 3225--3232 (2020).

\bibitem{Bo15}
H.~Bouchard, J.~de~Pooter, A.~Bielajew, and S.~Duane,
\newblock {Reference dosimetry in the presence of magnetic fields: conditions
  to validate Monte Carlo simulations},
\newblock Phys. Med. Biol. {\bf 60}, 6639 -- 6654 (2015).

\bibitem{Pa09a}
M.~Pacilio, N.~Lanconelli, S.~Lo~Meo, M.~Betti, L.~Montani, L.~Torres~Aroche,
  and M.~Coca~P{\'e}rez,
\newblock Differences among Monte Carlo codes in the calculations of voxel
  values for radionuclide targeted therapy and analysis of their impact on
  absorbed dose evaluations,
\newblock Medical physics {\bf 36}, 1543--1552 (2009).

\bibitem{Hi15}
E.~Hippel{\"a}inen, M.~Tenhunen, and A.~Sohlberg,
\newblock Fast voxel-level dosimetry for 177Lu labelled peptide treatments,
\newblock Physics in Medicine \& Biology {\bf 60}, 6685 (2015).

\bibitem{Bo99g}
W.~E. Bolch et~al.,
\newblock MIRD pamphlet no. 17: the dosimetry of nonuniform activity
  distributions—radionuclide S values at the voxel level,
\newblock Journal of Nuclear Medicine {\bf 40}, 11S--36S (1999).

\bibitem{Mi17}
N.~Miksys, E.~Vigneault, A.-G. Martin, L.~Beaulieu, and R.~M. Thomson,
\newblock Large-scale retrospective Monte Carlo dosimetric study for permanent
  implant prostate brachytherapy,
\newblock Int. J. Radiat. Oncol. Biol. Phys. {\bf 97}, 606--615 (2017).

\bibitem{Vi19}
E.~Vigneault, K.~Mbodji, D.~Carignan, A.-G. Martin, N.~Miksys, R.~M. Thomson,
  S.~Aubin, N.~Varfalvy, and L.~Beaulieu,
\newblock The association of intraprostatic calcifications and dosimetry
  parameters with biochemical control after permanent prostate implant,
\newblock Brachytherapy {\bf 18}, 787--792 (2019).

\bibitem{Ka99b}
I.~Kawrakow,
\newblock { Accurate condensed history Monte Carlo simulation of electron
  transport. II. Application to ion chamber response simulations},
\newblock Med. Phys. {\bf 27}, 499 -- 513 (2000).

\bibitem{Se02}
J.~P. Seuntjens, I.~Kawrakow, J.~Borg, F.~Hobeila, and D.~W.~O. Rogers,
\newblock {Calculated and measured air-kerma response of ionization chambers in
  low and medium energy photon beams},
\newblock in {\em Recent developments in accurate radiation dosimetry, Proc. of
  an Int'l Workshop}, edited by J.~P. Seuntjens and P.~Mobit, pages 69 -- 84,
  Medical Physics Publishing, Madison WI, 2002.

\bibitem{Mi16}
N.~Miksys,
\newblock {\em Advancements in Monte Carlo Dose Calculations for Prostate and
  Breast Permanent Implant Brachytherapy},
\newblock PhD thesis, Carleton University, 2016.

\end{thebibliography}

\end{document}